\PassOptionsToPackage{unicode}{hyperref}
\PassOptionsToPackage{hyphens}{url}
\documentclass[
]{article}
\usepackage{amsmath,amssymb}
\usepackage{iftex}
\ifPDFTeX
  \usepackage[T1]{fontenc}
  \usepackage[utf8]{inputenc}
  \usepackage{textcomp} 
\else 
  \usepackage{unicode-math} 
  \defaultfontfeatures{Scale=MatchLowercase}
  \defaultfontfeatures[\rmfamily]{Ligatures=TeX,Scale=1}
\fi
\usepackage{lmodern}
\ifPDFTeX\else
\fi
\IfFileExists{upquote.sty}{\usepackage{upquote}}{}
\IfFileExists{microtype.sty}{
  \usepackage[]{microtype}
  \UseMicrotypeSet[protrusion]{basicmath} 
}{}
\makeatletter
\@ifundefined{KOMAClassName}{
  \IfFileExists{parskip.sty}{%
    \usepackage{parskip}
  }{
    \setlength{\parindent}{0pt}
    \setlength{\parskip}{6pt plus 2pt minus 1pt}}
}{
  \KOMAoptions{parskip=half}}
\makeatother
\usepackage{xcolor}
\usepackage{longtable,booktabs,array}
\usepackage{calc} 
\usepackage{etoolbox}
\makeatletter
\patchcmd\longtable{\par}{\if@noskipsec\mbox{}\fi\par}{}{}
\makeatother
\IfFileExists{footnotehyper.sty}{\usepackage{footnotehyper}}{\usepackage{footnote}}
\makesavenoteenv{longtable}
\usepackage{graphicx}
\makeatletter
\def\maxwidth{\ifdim\Gin@nat@width>\linewidth\linewidth\else\Gin@nat@width\fi}
\def\maxheight{\ifdim\Gin@nat@height>\textheight\textheight\else\Gin@nat@height\fi}
\makeatother
\setkeys{Gin}{width=\maxwidth,height=\maxheight,keepaspectratio}
\makeatletter
\def\fps@figure{htbp}
\makeatother
\providecommand{\tightlist}{%
  \setlength{\itemsep}{0pt}\setlength{\parskip}{0pt}}
\newlength{\cslhangindent}
\newlength{\csllabelwidth}
\newlength{\cslentryspacingunit} 
\newenvironment{CSLReferences}[2] 
 {
  \setlength{\parindent}{0pt}
  \ifodd #1
  \let\oldpar\par
  \def\par{\hangindent=\cslhangindent\oldpar}
  \fi
  \setlength{\parskip}{#2\cslentryspacingunit}
 }%
 {}
\usepackage{calc}

\newcommand{\CSLLeftMargin}[1]{\parbox[t]{\csllabelwidth}{#1}}
\newcommand{\CSLRightInline}[1]{\parbox[t]{\linewidth - \csllabelwidth}{#1}\break}

\ifLuaTeX
  \usepackage{selnolig}  
\fi
\IfFileExists{bookmark.sty}{\usepackage{bookmark}}{\usepackage{hyperref}}
\IfFileExists{xurl.sty}{\usepackage{xurl}}{} 
\hypersetup{
  hidelinks,
  pdfcreator={LaTeX via pandoc}}

\author{}
\date{}

\begin{document}

\hypertarget{large-language-models-for-automated-prisma-2020-adherence-checking}{%
\section{Large language models for automated PRISMA 2020 adherence
checking}\label{large-language-models-for-automated-prisma-2020-adherence-checking}}

\textbf{Authors}\\
Yuki Kataoka, MD, MPH, DrPH\^{}1,2,3,4,5,6\^{}; Ryuhei So, MD, MPH,
PhD\textsuperscript{3,7,8}; Masahiro Banno, MD,
PhD\textsuperscript{3,9}; Yasushi Tsujimoto, MD, MPH,
PhD\textsuperscript{3,10,11}; Tomohiro Takayama, MD\textsuperscript{12};
Yosuke Yamagishi, MD/MSc\textsuperscript{13}; Takahiro Tsuge, PT,
MPH\textsuperscript{3,14,15}; Norio Yamamoto, MD,
PhD\textsuperscript{3,16}; Chiaki Suda, MD\textsuperscript{3,17}; Toshi
A. Furukawa, MD, PhD\textsuperscript{11}

\textbf{Affiliations}\\
\textsuperscript{1} Center for Postgraduate Clinical Training and Career
Development, Nagoya University Hospital, Nagoya, Aichi, Japan
\textsuperscript{2} Center for Medical Education, Graduate School of
Medicine, Nagoya University, Nagoya, Aichi, Japan\\
\textsuperscript{3} Scientific Research Works Peer Support Group
(SRWS-PSG), Osaka, Japan\\
\textsuperscript{4} Department of Internal Medicine, Kyoto Min-iren
Asukai Hospital, Kyoto, Japan\\
\textsuperscript{5} Department of Healthcare Epidemiology, Kyoto
University Graduate School of Medicine / School of Public Health, Kyoto,
Japan\\
\textsuperscript{6} Department of International and Community Oral
Health, Tohoku University Graduate School of Dentistry, Sendai, Miyagi,
Japan\\
\textsuperscript{7} Department of Psychiatry, Okayama Psychiatric
Medical Center, Okayama, Japan\\
\textsuperscript{8} CureApp, Inc., Tokyo, Japan\\
\textsuperscript{9} Department of Psychiatry, Seichiryo Hospital,
Nagoya, Japan\\
\textsuperscript{10} Oku medical clinic, Osaka, Japan\\
\textsuperscript{11} Department of Health Promotion and Human Behavior,
Kyoto University Graduate School of Medicine / School of Public Health,
Kyoto, Japan\\
\textsuperscript{12} Kyoto University Hospital, Kyoto, Japan\\
\textsuperscript{13} Division of Radiology and Biomedical Engineering,
Graduate School of Medicine, The University of Tokyo, Tokyo, Japan\\
\textsuperscript{14} Department of Rehabilitation, Kurashiki Medical
Centre, Kurashiki, Okayama, Japan\\
\textsuperscript{15} Department of Epidemiology, Graduate School of
Medicine, Dentistry, and Pharmaceutical Sciences, Okayama University,
Okayama, Japan\\
\textsuperscript{16} Department of Orthopedic Surgery, Minato Medical
Coop-Kyoritsu General Hospital, Nagoya, Aichi, Japan\\
\textsuperscript{17} Department of Public Health, Gunma University
Graduate School of Medicine, Gunma, Japan

\textbf{Corresponding author}\\
\emph{Correspondence}: Yuki Kataoka (youkiti@gmail.com)\\
Address: Center for Postgraduate Clinical Training and Career
Development, Nagoya University Hospital, 65, Tsurumai-cho, Showa-ku,
Nagoya-city, Aichi, Japan

\begin{center}\rule{0.5\linewidth}{0.5pt}\end{center}

\hypertarget{abstract}{%
\section{Abstract}\label{abstract}}

Evaluating adherence to PRISMA 2020 guideline remains a burden in the
peer review process. To address the lack of shareable benchmarks, we
constructed a copyright-aware benchmark of 108 Creative Commons-licensed
systematic reviews and evaluated ten large language models (LLMs) across
five input formats. In a development cohort, supplying structured PRISMA
2020 checklists (Markdown, JSON, XML, or plain text) yielded
78.7--79.7\% accuracy versus 45.21\% for manuscript-only input
(p\textless0.0001), with no differences between structured formats
(p\textgreater0.9). Across models, accuracy ranged from 70.6--82.8\%
with distinct sensitivity--specificity trade-offs, replicated in an
independent validation cohort. We then selected Qwen3-Max (a
high-sensitivity open-weight model) and extended evaluation to the full
dataset (n=120), achieving 95.1\% sensitivity and 49.3\% specificity.
Structured checklist provision substantially improves LLM-based PRISMA
assessment, though human expert verification remains essential before
editorial decisions.

\hypertarget{introduction}{%
\section{Introduction}\label{introduction}}

Transparent and complete reporting underpins credible evidence
synthesis. The Preferred Reporting Items for Systematic Reviews and
Meta-Analyses (PRISMA) 2020 statement was introduced to improve
reporting in systematic reviews (SRs) (1), yet recent cross-sectional
analyses revealed persistent gaps in adherence across fields (2--4).
Adherence checks are typically performed manually by reviewers and
editors, contributing to an already heavy peer-review workload (5--7).

Recent advances in LLMs suggest potential for assisting adherence
assessment in reporting guidelines (8--13). However, prompt formatting
substantially influences evaluation performance: even when assessing
identical content, input structure (e.g., Markdown vs.~JSON vs.~plain
text) can alter LLM accuracy by up to 40\% (14). No prior study has
systematically compared checklist input formats for PRISMA adherence
assessment, nor established shareable benchmarks with item-level labels
on full-text systematic reviews that respect copyright constraints (15).
These gaps limit evidence-based deployment of LLM-assisted adherence
checking in peer-review workflows.

This study systematically evaluates LLM performance in PRISMA 2020
adherence assessment. We compared checklist input formats (Markdown,
JSON, XML, plain text, control) to identify optimal representations,
then evaluated ten LLM models using the selected format, extended the
validation phase to twelve models by adding Gemini 3 Pro and GPT-5.1,
and examined item-level accuracy profiles for individual PRISMA
checklist items.

\hypertarget{results}{%
\section{Results}\label{results}}

\hypertarget{study-overview}{%
\subsection{Study overview}\label{study-overview}}

We report this study following the transparent reporting of a
multivariable model for individual prognosis or diagnosis (TRIPOD)-LLM
statement (16) (Supplemental table 1). Figure 1 shows the complete
three-phase methodological pipeline: parameter optimization, format and
model comparison, and validation across clinical domains.

\begin{figure}
\centering
\includegraphics{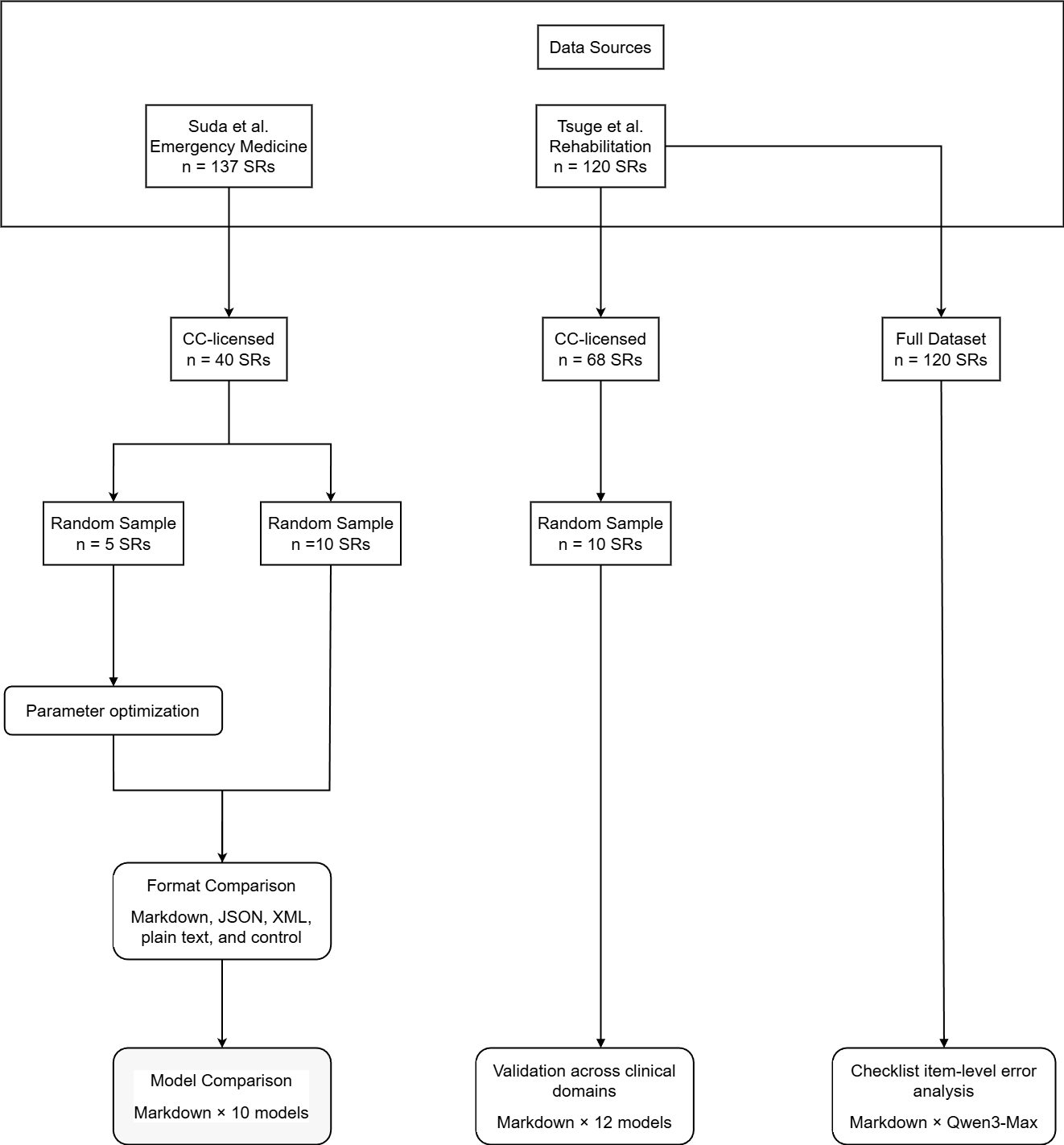}
\caption{Figure 1. Study overview and methodological pipeline}
\end{figure}

\hypertarget{dataset-characteristics}{%
\subsection{Dataset characteristics}\label{dataset-characteristics}}

From two source datasets---Suda et al.~(emergency medicine; 137 records)
(3) and Tsuge et al.~(rehabilitation; 120 records) (4)---we identified
40 articles (29.2\%) and 68 articles (56.7\%), respectively, with
Creative Commons licenses eligible for public release, totaling 108 SRs.
Across this dataset, after converting the full-text PDFs into a
structured text format, abstracts contained a median of 264 words
(interquartile range 238--295; Suda median 282; Tsuge median 253), and
main bodies contained a median of 4,674 words (interquartile range
3,920--5,801; Suda median 4,239; Tsuge median 5,100). Abstract checklist
adherence had a median of 5 of 12 items (interquartile range 4--6; Suda
median 6; Tsuge median 5). Main body checklist adherence had a median of
28 of 41 items (interquartile range 25--32; Suda median 28; Tsuge median
28).

From the Suda et al.~dataset, we randomly sampled 5 SRs for parameter
optimization and 10 different SRs for format and model comparison. From
the Tsuge et al.~dataset, we randomly sampled 10 SRs for validation
across clinical domains.

\hypertarget{parameter-optimization}{%
\subsection{Parameter optimization}\label{parameter-optimization}}

We conducted parameter optimization experiments using five randomly
sampled SRs from the Suda et al.~dataset. First, to verify that
evaluation order did not affect performance, we tested Claude Opus 4.1
using two prompt arrangements: checklist-first versus manuscript-first.
Reversing the order produced minimal change in overall accuracy (80.4\%
vs 79.6\%), sensitivity (91.7\% vs 88.1\%), and specificity (50.0\% vs
56.9\%). We therefore standardized all subsequent experiments to supply
manuscript text immediately before the checklist. For Claude Opus 4.1,
we optimized the thinking-token budget---the number of word counts
allocated for internal step-by-step reasoning before generating the
final output. We compared four budget levels (24 000, 28 000, 30 000,
and 31 000 tokens). At 24 000 tokens, accuracy was 79.0\% with
specificity 81.4\% and sensitivity 95.0\%. At 28 000 tokens, accuracy
was 82.9\% with specificity 87.1\% and sensitivity 91.9\%. At 30 000
tokens, accuracy was 76.1\% with specificity 82.9\% and sensitivity
87.6\%. At 31 000 tokens, accuracy was 80.5\% with specificity 83.4\%
and sensitivity 93.8\%. The 28 000-token setting achieved optimal
balance between accuracy and specificity. Based on these results, we
fixed Claude's thinking budget at 28 000 tokens for all manuscript
evaluations with both Opus 4.1 and Sonnet 4.5.

For GPT-5, we evaluated three reasoning effort levels (minimal, medium,
and high). Minimal and medium settings failed to return complete
structured outputs. We therefore standardized all GPT-5 experiments to
use the high reasoning setting to maintain reproducible evaluations.

\hypertarget{format-comparison}{%
\subsection{Format comparison}\label{format-comparison}}

Following parameter optimization, we expanded evaluation to ten
additional randomly sampled SRs from the Suda et al.~dataset to compare
checklist input formats. We evaluated five representations---Markdown,
JSON, XML, plain text, and control---across all ten LLM models using the
locked parameter settings. Here ``control'' denotes the condition with
no PRISMA checklist supplied (manuscript-only input).

Integrating the results from all 10 LLMs, structured formats
substantially outperformed the control across all metrics (Figure 2).
Mean accuracy ranged from 78.7\% to 79.7\% for structured formats versus
45.2\% for the control (Markdown: 79.2\%, 95\% CI 77.0--81.3). Similar
patterns emerged for sensitivity (structured: 86.4--88.0\%; control:
55.3\%) and specificity (structured: 66.8--68.2\%; control: 31.3\%).

Statistical analysis confirmed these differences. One-way ANOVA revealed
between-format variation in accuracy, sensitivity, and specificity
(p\textless0.0001). Each structured format significantly outperformed
the control (p\textless0.01), whereas differences among structured
formats were not statistically significant (p\textgreater0.9). Based on
these findings showing comparable performance across structured formats,
we selected Markdown format for subsequent model comparison and
validation phases due to its superior human readability.

\begin{figure}
\centering
\includegraphics{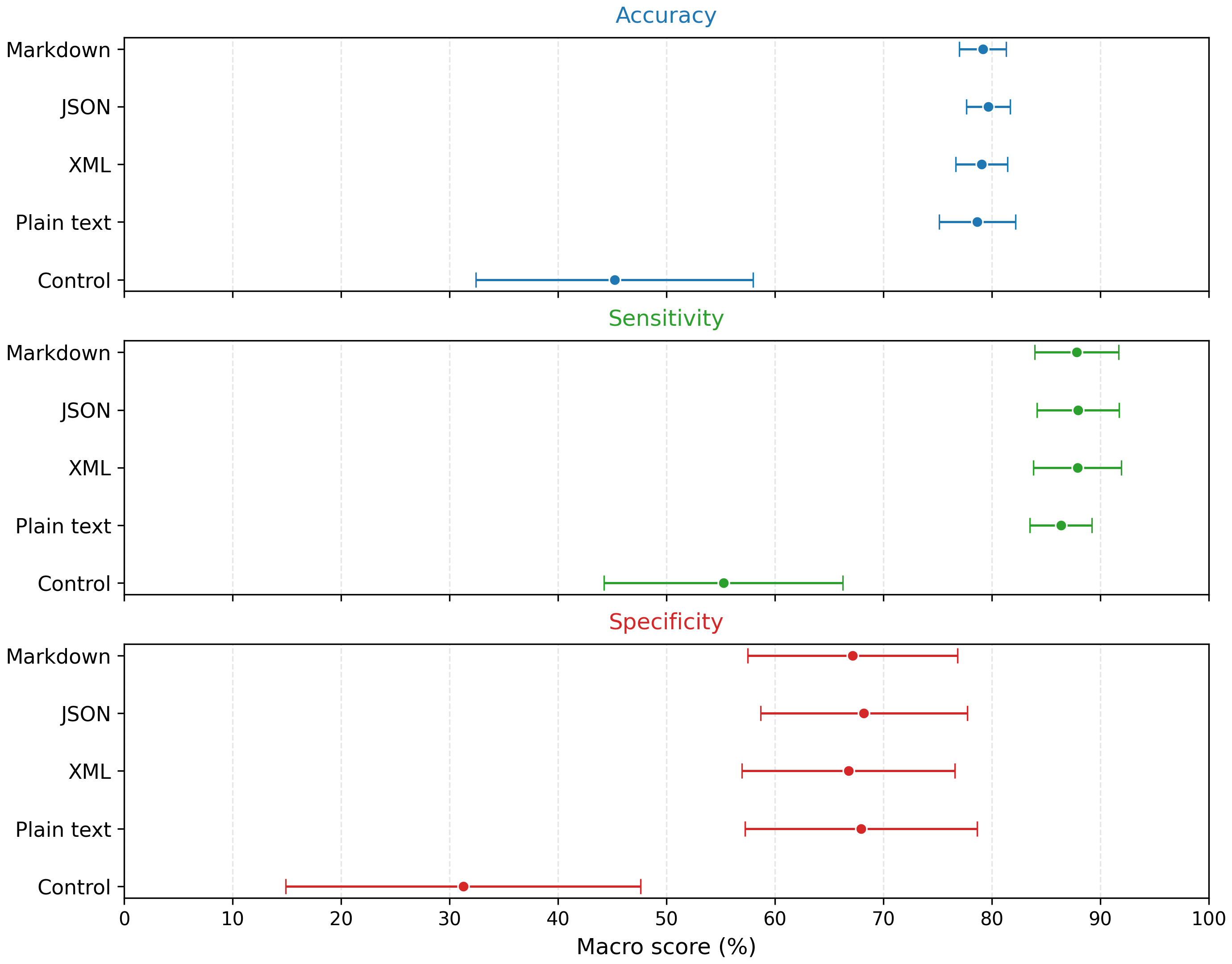}
\caption{Figure 2. Format-level macro metrics with 95\% CI (Suda et al.~dataset, n=10 SRs). Dot plots display mean accuracy, sensitivity, and specificity with 95\% confidence intervals for five checklist input formats evaluated on ten systematic reviews from the Suda et al.~dataset. Each data point represents the mean performance metric across all ten large language models for a given format.}
\end{figure}

\hypertarget{model-comparison}{%
\subsection{Model comparison}\label{model-comparison}}

Having established that structured formats outperformed unstructured
input and that differences among structured formats are minimal, we
conducted detailed model comparison using the selected Markdown format
on the same ten reviews. Accuracy varied across models, ranging from
70.6\% (GPT-4o) to 82.8\% (Grok-4-fast) (Figure 3). Models demonstrated
distinct sensitivity-specificity trade-offs. Sensitivity ranged from
77.2\% (GPT-5) to 97.3\% (GPT-4o), while specificity ranged from 33.9\%
(GPT-4o) to 85.7\% (GPT-5). High-sensitivity models (GPT-4o: 97.3\%;
Qwen3-235B: 94.9\%; Qwen3-Max: 94.6\%; Grok-4-fast: 90.3\%) generally
exhibited lower specificity (\textless71\%), whereas GPT-5 and Grok-4
maintained balanced profiles with both metrics \textgreater75\%.
Statistical analysis revealed between-model differences in all metrics
(ANOVA: accuracy p=0.0147, sensitivity and specificity both
p\textless0.0001). Post-hoc testing identified GPT-4o as having
significantly lower accuracy versus GPT-5 (adjusted p=0.018), higher
sensitivity than GPT-5 (adjusted p\textless0.05), and lower specificity
than GPT-5 (adjusted p\ensuremath{\leq}0.001). Processing time per SR
varied, ranging from 31.71 s (GPT-4o) to 282.4 s (GPT-5). Mean API cost
per review ranged from \$0.002 (GPT-OSS-120B) to \$0.60 (Claude Opus
4.1), with most models below \$0.20 (Table 2).

\textbf{Figure 3 \textbar{} Model-level macro metrics with 95\% CI
(Markdown input, Suda et al.~dataset, n=10).}

\begin{figure}
\centering
\includegraphics{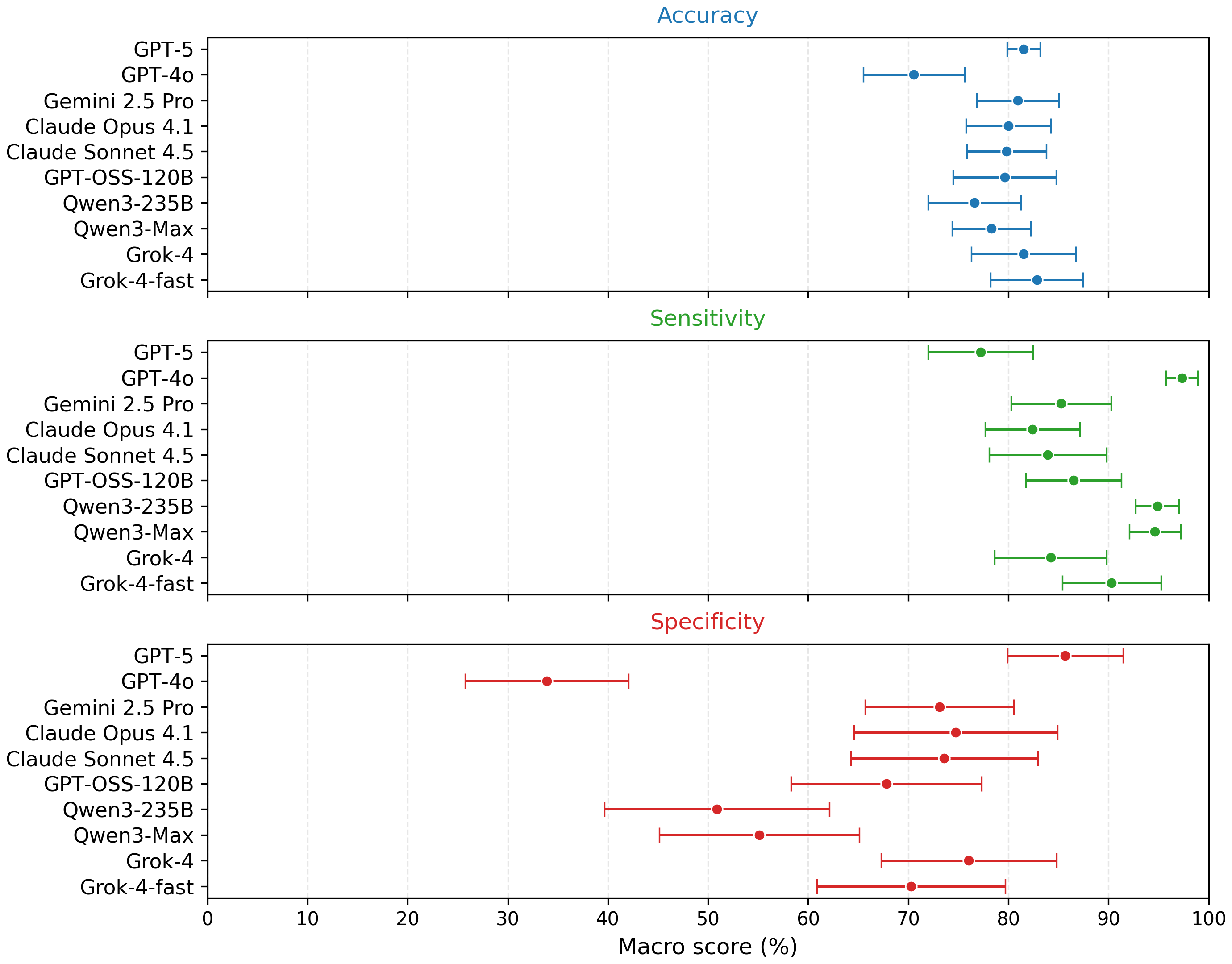}
\caption{Figure 3. Model-level macro metrics (Markdown input; values
shown)}
\end{figure}

Dot plots show mean accuracy, sensitivity, and specificity with 95\%
confidence intervals (horizontal error bars) for ten large language
models evaluated on ten systematic reviews from the Suda et al.~dataset
using Markdown checklist format.

\textbf{Table 2 \textbar{} Model-level computational performance metrics
(Markdown input; n=10 SRs)}

\begin{longtable}[]{@{}lrr@{}}
\toprule\noalign{}
Model & Mean time per SR (sec) & Cost per SR (USD) \\
\midrule\noalign{}
\endhead
\bottomrule\noalign{}
\endlastfoot
GPT-5 & 282.4 & 0.179 \\
GPT-4o & 31.7 & 0.047 \\
Gemini 2.5 Pro & 82.0 & 0.048 \\
Claude Opus 4.1 & 101.3 & 0.598 \\
Claude Sonnet 4.5 & 183.6 & 0.178 \\
GPT-OSS-120B & 45.0 & 0.002 \\
Qwen3-235B & 46.7 & 0.004 \\
Qwen3-Max & 47.8 & 0.030 \\
Grok-4 & 103.0 & 0.106 \\
Grok-4-fast & 36.2 & 0.005 \\
\end{longtable}

\hypertarget{validation-with-rehabilitation-dataset}{%
\subsection{Validation with rehabilitation
dataset}\label{validation-with-rehabilitation-dataset}}

To assess generalizability across clinical domains, we applied the same
evaluation approach (Markdown format with optimized parameters) to ten
randomly selected SRs from the Tsuge et al.~dataset. We evaluated twelve
models including two models released after the initial experiment phase
(Gemini 3 Pro and GPT-5.1). Five models exceeded 80\% accuracy: Grok-4
Fast (83.0\%, 95\% CI 79.6--86.0\%), Grok-4 (81.1\%, 95\% CI
77.6--84.2\%), Gemini 3 Pro (81.3\%, 95\% CI 77.8--84.4\%), GPT-OSS-120B
(80.9\%, 95\% CI 77.4--84.1\%), and GPT-5.1 (80.2\%, 95\% CI
76.6--83.4\%). Accuracy ranged from 68.5\% (95\% CI 64.4--72.3\%) for
GPT-4o to 83.0\% for Grok-4 Fast. The sensitivity-specificity trade-off
pattern persisted (Figure 4). High-sensitivity models GPT-4o (97.0\%,
95\% CI 94.3--98.4\%) and Qwen3-Max (96.0\%, 95\% CI 93.1--97.7\%)
maintained this performance profile but exhibited low specificity
(32.2\%, 95\% CI 26.5--38.4\% and 55.0\%, 95\% CI 48.5--61.2\%).

\begin{figure}
\centering
\includegraphics{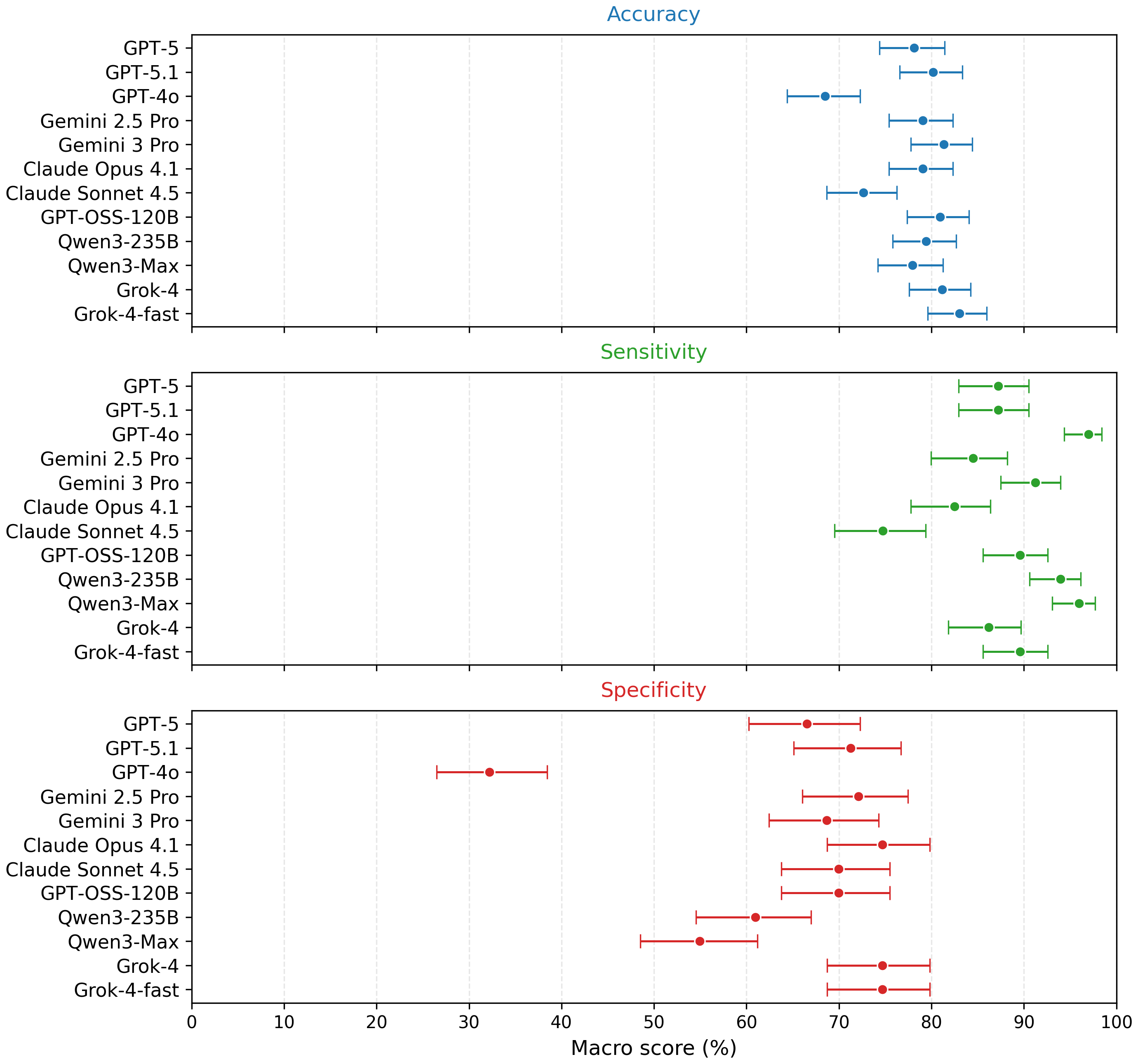}
\caption{Figure 4. Validation-phase model-level metrics with 95\% CI (Markdown input, Tsuge et al.~dataset, n=10). Dot plots show mean accuracy, sensitivity, and specificity with 95\% confidence intervals (horizontal error bars) for twelve large language models evaluated on ten systematic reviews from the Tsuge et al.~dataset using Markdown checklist format. Performance patterns observed in format and model comparison phase were confirmed in this independent validation cohort.}
\end{figure}

Processing time per SR varied, ranging from 27.0 s (Grok-4 Fast) to
207.8 s (Claude Sonnet 4.5). Mean API cost per review ranged from
\$0.002 (GPT-OSS-120B) to \$0.562 (Claude Opus 4.1) (Table 3).

\textbf{Table 3 \textbar{} Validation-phase computational performance
metrics (Markdown input; n=10 SRs)}

\begin{longtable}[]{@{}lrr@{}}
\toprule\noalign{}
Model & Mean time per SR (sec) & Cost per SR (USD) \\
\midrule\noalign{}
\endhead
\bottomrule\noalign{}
\endlastfoot
GPT-5 & 185.4 & 0.181 \\
GPT-5.1 & 140.4 & 0.159 \\
GPT-4o & 31.8 & 0.043 \\
Gemini 2.5 Pro & 76.4 & 0.045 \\
Gemini 3 Pro & 113.0 & 0.054 \\
Claude Opus 4.1 & 122.4 & 0.562 \\
Claude Sonnet 4.5 & 207.8 & 0.163 \\
GPT-OSS-120B & 42.8 & 0.002 \\
Qwen3-235B & 53.7 & 0.003 \\
Qwen3-Max & 56.1 & 0.027 \\
Grok-4 & 117.8 & 0.111 \\
Grok-4 Fast & 27.0 & 0.005 \\
\end{longtable}

Runtime and API cost for the Tsuge rehabilitation validation set
(Markdown checklist input). Cost reflects total API spend per systematic
review (USD, per 1M-token pricing converted directly from provider
invoices).

\hypertarget{item-level-error-profile}{%
\subsection{Item-level error profile}\label{item-level-error-profile}}

To characterize systematic error patterns across PRISMA items, we
conducted an extended evaluation using Qwen3-Max with Markdown checklist
on the complete Tsuge rehabilitation dataset (n=120 SRs, including both
Creative Commons-licensed and non-licensed articles). Overall
performance on the full dataset showed 78.2\% accuracy, with 95.1\%
sensitivity and 49.3\% specificity. Main-body items achieved 80.1\%
accuracy, while abstract items showed 71.8\% accuracy. Performance was
comparable between Creative Commons-licensed articles (n=68; accuracy
78.1\%, sensitivity 95.1\%, specificity 49.0\%) and non-licensed
articles (n=52; accuracy 78.5\%, sensitivity 95.1\%, specificity
50.0\%). Item-level analysis revealed systematic error patterns (Table
4a, 4b). The highest FN proportions were observed for five items:
main\_24c (availability of data/code; 86.2\%), main\_27 (funding
sources; 60.4\%), main\_7 (information sources; 53.2\%),
abstract\_item\_11 (interpretation; 42.9\%), and main\_24b (registration
availability; 20.0\%). Fourteen items showed 100\% FP proportions.

\textbf{Table 4a \textbar{} False negative-prone items in item-level
error profile (n=120 SRs)}

\begin{longtable}[]{@{}
  >{\raggedright\arraybackslash}p{(\columnwidth - 8\tabcolsep) * \real{0.1579}}
  >{\raggedleft\arraybackslash}p{(\columnwidth - 8\tabcolsep) * \real{0.2105}}
  >{\raggedleft\arraybackslash}p{(\columnwidth - 8\tabcolsep) * \real{0.2105}}
  >{\raggedleft\arraybackslash}p{(\columnwidth - 8\tabcolsep) * \real{0.2105}}
  >{\raggedleft\arraybackslash}p{(\columnwidth - 8\tabcolsep) * \real{0.2105}}@{}}
\toprule\noalign{}
\begin{minipage}[b]{\linewidth}\raggedright
Item ID (meaning)
\end{minipage} & \begin{minipage}[b]{\linewidth}\raggedleft
TP
\end{minipage} & \begin{minipage}[b]{\linewidth}\raggedleft
FN
\end{minipage} & \begin{minipage}[b]{\linewidth}\raggedleft
Total Positive
\end{minipage} & \begin{minipage}[b]{\linewidth}\raggedleft
FN Proportion
\end{minipage} \\
\midrule\noalign{}
\endhead
\bottomrule\noalign{}
\endlastfoot
main 24c (Registration and protocol -- amendments) & 8 & 50 & 58 &
86.2\% \\
main 27 (Availability of data) & 36 & 55 & 91 & 60.4\% \\
main 7 (Search strategy) & 29 & 33 & 62 & 53.2\% \\
abstract item 11 (Funding) & 4 & 3 & 7 & 42.9\% \\
main 24b (Registration and protocol -- access) & 12 & 3 & 15 & 20.0\% \\
main 16b (Study selection -- excluded studies) & 17 & 2 & 19 & 10.5\% \\
main 13e (Synthesis methods -- heterogeneity) & 50 & 5 & 55 & 9.1\% \\
main 21 (Reporting biases) & 54 & 3 & 57 & 5.3\% \\
main 26 (Competing interests) & 108 & 6 & 114 & 5.3\% \\
main 14 (Reporting bias assessment) & 63 & 3 & 66 & 4.5\% \\
abstract item 12 (Registration) & 22 & 1 & 23 & 4.3\% \\
main 8 (Selection process) & 102 & 3 & 105 & 2.9\% \\
main 25 (Support) & 102 & 3 & 105 & 2.9\% \\
main 22 (Certainty of evidence) & 42 & 1 & 43 & 2.3\% \\
main 15 (Certainty assessment) & 43 & 1 & 44 & 2.3\% \\
main 20d (Results of syntheses -- sensitivity analyses) & 44 & 1 & 45 &
2.2\% \\
main 9 (Data collection process) & 90 & 2 & 92 & 2.2\% \\
main 24a (Registration and protocol -- registration) & 90 & 2 & 92 &
2.2\% \\
main 13f (Synthesis methods -- sensitivity analyses) & 48 & 1 & 49 &
2.0\% \\
main 20c (Results of syntheses -- heterogeneity) & 48 & 1 & 49 &
2.0\% \\
main 23c (Discussion -- limitations) & 99 & 1 & 100 & 1.0\% \\
main 18 (Risk-of-bias in studies) & 117 & 1 & 118 & 0.8\% \\
\end{longtable}

Items with at least one false negative case, sorted by FN rate
(descending). TP: true positives; FN: false negatives; Total Positive:
TP + FN (cases where human annotation = Yes).

\textbf{Table 4b \textbar{} False positive-prone items in item-level
error profile (n=120 SRs)}

\begin{longtable}[]{@{}
  >{\raggedright\arraybackslash}p{(\columnwidth - 8\tabcolsep) * \real{0.1579}}
  >{\raggedleft\arraybackslash}p{(\columnwidth - 8\tabcolsep) * \real{0.2105}}
  >{\raggedleft\arraybackslash}p{(\columnwidth - 8\tabcolsep) * \real{0.2105}}
  >{\raggedleft\arraybackslash}p{(\columnwidth - 8\tabcolsep) * \real{0.2105}}
  >{\raggedleft\arraybackslash}p{(\columnwidth - 8\tabcolsep) * \real{0.2105}}@{}}
\toprule\noalign{}
\begin{minipage}[b]{\linewidth}\raggedright
Item ID (meaning)
\end{minipage} & \begin{minipage}[b]{\linewidth}\raggedleft
TN
\end{minipage} & \begin{minipage}[b]{\linewidth}\raggedleft
FP
\end{minipage} & \begin{minipage}[b]{\linewidth}\raggedleft
Total Negative
\end{minipage} & \begin{minipage}[b]{\linewidth}\raggedleft
FP Proportion
\end{minipage} \\
\midrule\noalign{}
\endhead
\bottomrule\noalign{}
\endlastfoot
main 5 (Eligibility criteria) & 0 & 10 & 10 & 100.0\% \\
main 10a (Data items -- outcomes) & 0 & 68 & 68 & 100.0\% \\
main 10b (Data items -- other variables) & 0 & 75 & 75 & 100.0\% \\
main 12 (Effect measures) & 0 & 3 & 3 & 100.0\% \\
main 13a (Synthesis methods -- studies to synthesis) & 0 & 6 & 6 &
100.0\% \\
main 13c (Synthesis methods -- tabulation/visualization) & 0 & 92 & 92 &
100.0\% \\
main 13d (Synthesis methods -- synthesis approach) & 0 & 38 & 38 &
100.0\% \\
main 17 (Study characteristics) & 0 & 1 & 1 & 100.0\% \\
main 19 (Results of individual studies) & 0 & 7 & 7 & 100.0\% \\
main 20b (Results of syntheses -- statistical results) & 0 & 3 & 3 &
100.0\% \\
main 23b (Discussion -- evidence limitations) & 0 & 7 & 7 & 100.0\% \\
main 23c (Discussion -- process limitations) & 0 & 20 & 20 & 100.0\% \\
main 23d (Discussion -- practice/policy/research implications) & 0 & 2 &
2 & 100.0\% \\
abstract item 10 (Interpretation) & 0 & 6 & 6 & 100.0\% \\
abstract item 6 (Synthesis of results) & 3 & 90 & 93 & 96.8\% \\
main 11 (Study risk of bias assessment) & 1 & 27 & 28 & 96.4\% \\
abstract item 7 (Included studies) & 5 & 93 & 98 & 94.9\% \\
main 6 (Information sources) & 3 & 49 & 52 & 94.2\% \\
main 8 (Selection process) & 1 & 14 & 15 & 93.3\% \\
main 1 (Title) & 1 & 13 & 14 & 92.9\% \\
main 9 (Data collection process) & 2 & 26 & 28 & 92.9\% \\
abstract item 1 (Title) & 1 & 13 & 14 & 92.9\% \\
main 20a (Results of syntheses -- summary) & 1 & 9 & 10 & 90.0\% \\
abstract item 8 (Synthesis of results) & 6 & 26 & 32 & 81.2\% \\
main 13b (Synthesis methods -- data preparation) & 16 & 63 & 79 &
79.7\% \\
abstract item 4 (Information sources) & 36 & 57 & 93 & 61.3\% \\
abstract item 3 (Eligibility criteria) & 46 & 71 & 117 & 60.7\% \\
main 18 (Risk of bias in studies) & 1 & 1 & 2 & 50.0\% \\
main 24b (Registration and protocol -- access) & 56 & 49 & 105 &
46.7\% \\
main 13e (Synthesis methods -- heterogeneity) & 39 & 26 & 65 & 40.0\% \\
main 20c (Results of syntheses -- heterogeneity) & 43 & 28 & 71 &
39.4\% \\
main 21 (Reporting biases) & 40 & 23 & 63 & 36.5\% \\
abstract item 9 (Limitations) & 62 & 33 & 95 & 34.7\% \\
main 7 (Search strategy) & 39 & 19 & 58 & 32.8\% \\
main 14 (Reporting bias assessment) & 38 & 16 & 54 & 29.6\% \\
main 20d (Results of syntheses -- sensitivity analysis) & 55 & 20 & 75 &
26.7\% \\
main 16b (Study selection -- excluded studies) & 78 & 23 & 101 &
22.8\% \\
main 13f (Synthesis methods -- sensitivity analysis) & 55 & 16 & 71 &
22.5\% \\
abstract item 5 (Risk of bias) & 59 & 12 & 71 & 16.9\% \\
main 26 (Competing interests) & 5 & 1 & 6 & 16.7\% \\
main 27 (Availability of data) & 25 & 4 & 29 & 13.8\% \\
main 25 (Support) & 13 & 2 & 15 & 13.3\% \\
main 24c (Registration and protocol -- amendments) & 58 & 4 & 62 &
6.5\% \\
main 15 (Certainty assessment) & 73 & 3 & 76 & 3.9\% \\
main 22 (Certainty of evidence) & 75 & 2 & 77 & 2.6\% \\
abstract item 11 (Funding) & 112 & 1 & 113 & 0.9\% \\
\end{longtable}

Items with at least one false positive case, sorted by FP rate
(descending). TN: true negatives; FP: false positives; Total Negative:
TN + FP (cases where human annotation = No).

\hypertarget{discussion}{%
\section{Discussion}\label{discussion}}

This study made a copyright-aware, shareable benchmark and the first
systematic evaluation of input format effects on LLM-based PRISMA 2020
adherence assessment. Providing structured checklists---regardless of
specific format (Markdown, JSON, XML, or plain text)---substantially
improved evaluation performance compared with manuscript-only input
without checklist guidance. Among structured formats, performance
differences were minimal. Model-level performance varied considerably,
with distinct sensitivity-specificity trade-offs. These performance
patterns persisted in a different clinical domain in the validation
phase. Item-level error analysis revealed systematic misclassification
patterns, with false negatives concentrated in metadata items such as
data availability and funding sources, and false positives distributed
across methodological detail items including process descriptions and
analysis methods.

The observed accuracy levels align with the complexity of PRISMA 2020
evaluation as a judgment task. Our results are comparable to studies
evaluating other reporting guidelines that require judgment: CONSORT
guideline evaluation achieving 80-90\% accuracy with high specificity
but lower sensitivity (12), and CONSORT/SPIRIT assessment with
reasoning-capable LLMs achieving approximately 85\% accuracy (10).
Previous PRISMA 2020 evaluation using LLMs showed systematic
overestimation of adherence by 23-30\% compared with human experts (13).
These studies typically reported performance for a single model
performance. In contrast, our multi-model evaluation revealed distinct
sensitivity-specificity trade-offs---from high-sensitivity models for
screening to balanced profiles for comprehensive evaluation. These
findings imply the importance of systematic model comparison before
deployment, as performance characteristics vary substantially across
models for the same task.

While current performance remains insufficient for fully automated human
replacement, our findings suggest that LLMs can effectively assist
PRISMA 2020 compliance evaluation within a human-in-the-loop
framework(7). For practical implementation, model selection should
consider the intended use case. High-sensitivity models (GPT-4o, Qwen3
variants) may be preferable for initial screening to minimize missed
items. The identified false negative-prone items (data availability,
funding sources, information sources) warrant particular human attention
during review. Since models can provide structured rationales alongside
binary decisions, reviewers can efficiently verify flagged items,
potentially reducing overall review burden compared with manual
assessment. Among open-weight alternatives, Qwen3 would provide
cost-effectiveness with acceptable performance.

The World Association of Medical Editors recommends that all parties in
the peer review process explicitly disclose LLMs usage (17). Following
this recommendation, we propose that using LLMs to assess unpublished
manuscripts is acceptable when security measures are comparable to those
of tools researchers routinely use, such as free email or general cloud
services. Specifically, we judge the usage to be acceptable under the
following conditions: (1) when AI is used, all parties (authors,
reviewers, editors) mutually disclose this usage within the peer review
process and transparently report it in the final manuscript, (2) the
platform meets baseline cloud-storage protections (encryption at rest
and in transit, access controls)(18), and (3) model training on user
inputs is disabled (via chat UI settings or API). Under these
conditions, we consider LLM support permissible for draft evaluation
without requiring strictly local model hosting.

Regarding the development of reporting guidelines more broadly, our
findings suggest that guideline developers should provide structured
machine-readable formats alongside traditional table documentation.
Given comparable LLM performance across structured formats, Markdown
offers particular advantages as the single format that maintains both
human readability for manual verification and machine parseability for
automated evaluation (19).

Several limitations warrant consideration. First, our domain scope is
limited to emergency medicine and rehabilitation, which may affect
generalizability to other medical specialties. To advance the field,
researchers should release human-annotated datasets for reporting
guideline evaluation through controlled-access mechanisms to prevent
benchmark contamination in future model training(20). Second, human
reference labels may retain residual inconsistencies despite
standardized annotation protocols, potentially affecting the accuracy of
performance metrics. Third, current performance remains insufficient for
fully automated evaluation. Future work should explore other reasoning
strategies such as ensemble approaches and promptings(21), as well as
systematic evaluation of emerging state-of-the-art models. Our public
benchmark enables systematic reassessment of these strategies and
emerging models. Fourth, deployment poses challenges for users lacking
API or coding expertise. However, our prompt enables evaluation of
individual papers using ChatGPT (GPT-4o) without programmatic setup
(22).

In conclusion, our findings demonstrate that researchers, peer
reviewers, and editors can use LLMs with structured checklist to help
identifying unreported PRISMA checklist items. However, given the
observed proportion of false positives, human expert review remains
essential for verification of flagged items before making decisions. The
field requires continued advancement through public dataset releases,
systematic evaluation of emerging models, and development of
human-in-the-loop workflows that balance automation efficiency with
expert oversight.

\hypertarget{methods}{%
\section{Methods}\label{methods}}

\hypertarget{study-design-and-datasets}{%
\subsection{Study design and datasets}\label{study-design-and-datasets}}

We conducted a four-phase study to quantify how input representation
affects LLM assessment of PRISMA 2020 checklist adherence. The unit of
analysis was the PRISMA item decision (Yes/No) at the level of each SR.
We used existing, item-level reference annotations from published
studies in emergency medicine and rehabilitation (3,4) and evaluated
multiple LLMs under harmonized prompt and decoding settings.

\begin{itemize}
\tightlist
\item
  Parameter optimization: 5 randomly sampled SRs from the Suda et
  al.~dataset (emergency medicine) (3).
\item
  Format and model comparison: 10 different randomly sampled SRs from
  the Suda et al.~dataset (emergency medicine) (3).
\item
  Validation: 10 randomly sampled SRs from the Tsuge et al.~dataset
  (rehabilitation) (4).
\item
  Item-level error characterization: complete Tsuge et al.~dataset
  (n=120 SRs) evaluated with Qwen3-Max using Markdown format to
  investigate systematic misclassification patterns across PRISMA items.
\end{itemize}

\hypertarget{data-preparation-and-licensing}{%
\subsection{Data preparation and
licensing}\label{data-preparation-and-licensing}}

\hypertarget{full-text-acquisition-and-conversion}{%
\subsubsection{Full-text acquisition and
conversion}\label{full-text-acquisition-and-conversion}}

We collected full-text PDFs and appendices from publisher websites or
indexes and converted them into structured, machine-readable JSON files.
We used the Adobe PDF Services API to extract document structure, plain
text, tables, and renditions of tables/figures per SR (23). Non-PDF
source files (for example, DOCX) were converted to PDF before
extraction.

\hypertarget{license-identification-and-filtering}{%
\subsubsection{License identification and
filtering}\label{license-identification-and-filtering}}

To construct a shareable benchmark while respecting copyright
constraints, we identified CC-licensed articles from two source
datasets: Suda et al.~(emergency medicine; 137 records) (3) and Tsuge et
al.~(rehabilitation; 240 records) (4).

We automatically mapped article license statements to CC variants using
pattern matching, including keywords such as ``noncommercial,'' ``no
derivatives,'' ``share alike,'' ``attribution,'' and CC codes (e.g.,
``cc-by-nc-nd,'' ``cc-by-nc,'' ``cc-by-sa,'' ``cc-by,'' ``cc0''). For
public release, we included records under any Creative Commons license
(BY, BY-SA, BY-NC, BY-NC-ND) and excluded records with unknown or no
license. This filtering retained 108 Creative Commons-licensed articles
across the two cohorts (Suda: 40/137, 29.2\%; Tsuge PRISMA: 68/120,
56.7\%).

\hypertarget{sampling-strategy}{%
\subsubsection{Sampling strategy}\label{sampling-strategy}}

From the license-eligible articles, we randomly sampled papers for three
experimental phases: parameter optimization (n=5 from emergency
medicine), format and model comparison (n=10 from emergency medicine),
and validation (n=10 from rehabilitation).

\hypertarget{checklist-formats}{%
\subsection{Checklist formats}\label{checklist-formats}}

One investigator created a structured PRISMA 2020 checklist in JSON
format based on the original guideline (1), and another investigator
confirmed its accuracy. We then converted the JSON checklist to
Markdown, XML, and plain text representations. A control condition with
no PRISMA checklist supplied was also evaluated. These standardized
checklist files were embedded directly into prompts for model
evaluation.

\hypertarget{reference-standard}{%
\subsubsection{Reference standard}\label{reference-standard}}

As the reference standard, we used PRISMA 2020 checklist adherence
evaluated by two independent human investigators for each SR. Any
disagreements were resolved through discussion, with a third reviewer
consulted when necessary.

\hypertarget{llm-selection-and-api-access}{%
\subsection{LLM selection and API
access}\label{llm-selection-and-api-access}}

We evaluated ten state-of-the-art LLMs via their providers' APIs between
September and November 2025. Models accessed through direct provider
APIs included OpenAI GPT-5 and GPT-4o, Google Gemini 2.5 Pro, and
Anthropic Claude Opus 4.1 and Sonnet 4.5. Additional models were
accessed via OpenRouter: OpenAI GPT-OSS-120B, Alibaba Qwen3-235B and
Qwen3-Max, and xAI Grok-4 and Grok-4-fast.

\hypertarget{structured-outputs}{%
\subsection{Structured outputs}\label{structured-outputs}}

To avoid free-text variability and ensure item-level comparability, we
constrained models to return structured responses using each provider's
native structured-output mechanism. For each article, models produced a
JSON object indexed by PRISMA item IDs, with a binary decision (yes/no)
and a brief rationale per item.

\hypertarget{parameter-optimization-1}{%
\subsection{Parameter optimization}\label{parameter-optimization-1}}

To establish reproducible evaluation settings, we set temperature (a
parameter controlling output randomness, where 0.0 produces
deterministic responses) at 0.0 where provider APIs supported explicit
control to minimize randomness. We then conducted parameter optimization
experiments using five randomly sampled SRs from the Suda et al.~dataset
to determine optimal model-specific settings. Using Claude Opus 4.1 with
simple Markdown checklist format, we tested whether prompt ordering
(manuscript-first vs.~checklist-first) affected performance. We also
swept Claude's thinking-token budget across 24,000, 28,000, 30,000, and
31,000 tokens to identify the optimal balance between performance and
cost. For GPT-5, we tested reasoning effort levels (minimal, medium,
high) to determine stable configurations for structured output
generation.

Based on these experiments, we locked parameter settings for all
subsequent phases, standardizing prompt order as manuscript text
followed by checklist, Claude thinking budget at 28,000 tokens for both
Opus 4.1 and Sonnet 4.5, and GPT-5 reasoning at high effort. Each model
received a standardized prompt that presented the manuscript text
followed by the PRISMA 2020 checklist, instructing the model to evaluate
each checklist item as ``Yes'' or ``No'' with a brief rationale. The
control condition received only the manuscript text without the
checklist.

We minimized randomness while respecting each model's capabilities.
Where a temperature control was available, we set temperature=0.0. When
a model did not expose a temperature setting, we used the provider's
default setting. We did not add penalty terms or custom stop rules.
Output length limits followed each provider's cap. Model-specific locked
parameters were: - OpenAI GPT-5 (\texttt{gpt-5-2025-08-07}): no
temperature setting; run with concise output (verbosity=low) and high
reasoning. - OpenAI GPT-4o (\texttt{gpt-4o-latest}): temperature=0.0. -
Google Gemini 2.5 Pro (\texttt{gemini-2.5-pro}): temperature=0.0 and
top-p=1.0. - Anthropic Claude Opus 4.1
(\texttt{claude-opus-4-1-20250805}) and Sonnet 4.5
(\texttt{claude-sonnet-4-5-20250929}): thinking trace enabled with a
28,000-token budget. - OpenAI GPT-OSS-120B
(\texttt{openai/gpt-oss-120b}): temperature=0.0. - Alibaba Qwen3-235B
(\texttt{qwen/qwen3-235b-a22b-2507}) and Qwen3-Max
(\texttt{qwen/qwen3-max}): temperature=0.0. - xAI Grok-4
(\texttt{x-ai/grok-4-0709}) and Grok-4-fast (\texttt{x-ai/grok-4-fast}):
temperature=0.0.

\hypertarget{format-and-model-comparison}{%
\subsection{Format and model
comparison}\label{format-and-model-comparison}}

Following parameter optimization, we evaluated ten randomly sampled SRs
from the Suda et al.~dataset across five input formats (Markdown, JSON,
XML, plain text, control) and ten LLM models using the locked parameter
settings. We set the sample size at ten following precedent from prior
studies (24,25). This phase aimed to identify the optimal checklist
format by comparing performance across different structured
representations. Based on these findings showing comparable performance
across structured formats, we selected Markdown format for validation
due to its superior human readability.

\hypertarget{validation}{%
\subsection{Validation}\label{validation}}

After the development stage, we locked the prompt format, parameter, and
analysis code. The validation phase applied the locked methodological
pipeline to ten randomly selected SRs from the Tsuge et al.~dataset. We
additionally evaluated two models released after the initial experiment
phase: - OpenAI GPT-5.1 (\texttt{gpt-5.1-2025-11-13}): no temperature
setting; run with concise output (verbosity=low) and high reasoning. -
Google Gemini 3 Pro (\texttt{gemini-3-pro-preview}): temperature=1.0 and
thinkingLevel=``HIGH''.

\hypertarget{item-level-error-characterization}{%
\subsection{Item-level error
characterization}\label{item-level-error-characterization}}

To investigate systematic misclassification patterns, we evaluated the
complete Tsuge rehabilitation dataset (n=120 SRs, including 68 Creative
Commons-licensed and 52 non-licensed articles) using Qwen3-Max with the
locked Markdown format.

\hypertarget{outcomes}{%
\subsection{Outcomes}\label{outcomes}}

The primary outcome was item-level diagnostic performance for PRISMA
2020 adherence, summarized per article and then aggregated across
articles. We report mean accuracy, sensitivity, and specificity across
items, treating each item decision as a binary classification against
the human-annotated reference. Secondary outcomes included computational
characteristics relevant to deployability: per-paper token usage (token
means similar to word counts), latency (request-to-response wall-clock
time), and monetary cost estimated from token usage.

\hypertarget{statistical-analysis}{%
\subsection{Statistical analysis}\label{statistical-analysis}}

We summarized model performance using descriptive statistics at the
PRISMA item level. For each model--format--paper combination, AI
decisions were compared with human reference labels to obtain true
positives (TP), true negatives (TN), false positives (FP), and false
negatives (FN). From these counts we computed accuracy = (TP+TN)/N,
sensitivity (sensitivity) = TP/(TP+FN), and specificity = TN/(TN+FP),
where N is the number of comparable items with a human annotation.
Item-level denominators included all items marked comparable by the
evaluation pipeline, and counts were pooled across the 10 format and
model comparison papers and, separately, the 10 validation papers.

For proportions (accuracy, sensitivity, specificity), 95\% confidence
intervals were calculated using the Wilson score interval.

We compared performance metrics across the five input formats and across
the ten LLM models using one-way analysis of variance (ANOVA). For
format comparison, pairwise comparisons between each structured format
and the control used Welch's t-test with Bonferroni correction for
multiple testing. For model comparison, post-hoc pairwise comparisons
used Welch's t-test with Bonferroni correction. All statistical tests
were two-sided at \ensuremath{\alpha} = 0.05. All analyses used Python
3.12.3 with numpy 2.3.2, pandas 2.3.2, SciPy 1.16.1, matplotlib 3.10.6,
and scikit-learn 1.7.1.

\hypertarget{ethical-consideration}{%
\section{Ethical consideration}\label{ethical-consideration}}

This study analyzed only publicly available, published SR articles and
their de-identified supplementary materials. This work does not
constitute human participants research and falls outside the scope of
ethics committee review. Therefore, ethics approval and informed consent
were not required.

\hypertarget{data-and-code-availability}{%
\section{Data and code availability}\label{data-and-code-availability}}

Datasets are available at Zenodo
(https://doi.org/10.5281/zenodo.17547700). Source codes are available at
GitHub (https://github.com/youkiti/PRISMA-AI-Share).

\hypertarget{acknowledgements}{%
\section{Acknowledgements}\label{acknowledgements}}

The application programming interface fee was supported by a JSPS
Grant-in-Aid for Scientific Research (Grant No.~25K13585) provided to
Y.K. The funders had no role in study design, data collection and
analysis, decision to publish, or preparation of the manuscript.

\hypertarget{use-of-ai-tools}{%
\section{Use of AI tools}\label{use-of-ai-tools}}

We used OpenAI Codex and Anthropic Claude Code to draft the code, then
checked and validated to finalize the analysis. We used Claude 4.5
Sonnet for English editing. All authors reviewed and edited the final
manuscript.

\hypertarget{author-contributions}{%
\section{Author contributions}\label{author-contributions}}

Y.K.: conceptualization, methodology, software, formal analysis,
investigation, data curation, writing---original draft, visualization,
project administration, funding acquisition. R.S.: conceptualization,
methodology, validation, writing---review \& editing. M.B.:
conceptualization, methodology, validation, writing---review \& editing.
Y.T.: conceptualization, methodology, validation, writing---review \&
editing. T.T. (Takayama): investigation, data curation, writing---review
\& editing. Y.Y.: investigation, data curation, resources,
writing---review \& editing. T.T. (Tsuge): investigation, data curation,
validation, writing---review \& editing. N.Y.: conceptualization,
validation, writing---review \& editing. C.S.: investigation, data
curation, validation, writing---review \& editing. T.A.F.:
conceptualization, methodology, supervision, writing---review \&
editing. All authors gave final approval of the version to be published
and agreed to be accountable for all aspects of this work.

\hypertarget{competing-interests}{%
\section{Competing interests}\label{competing-interests}}

R.S. reports grants from the Osake-no-Kagaku Foundation, speaker's
honoraria from Otsuka Pharmaceutical Co., Ltd., Nippon Shinyaku Co.,
Ltd., and Takeda Pharmaceutical Co., Ltd., outside the submitted work.
T.A.F. has patents 2020-548587 and 2022-082495 pending, and intellectual
properties for Kokoro-app licensed to Mitsubishi-Tanabe. The remaining
authors declare no competing interests.

\hypertarget{references}{%
\section{References}\label{references}}

\hypertarget{refs}{}
\begin{CSLReferences}{0}{0}
\leavevmode\vadjust pre{\hypertarget{ref-Page2021-vy}{}}%
\CSLLeftMargin{1. }%
\CSLRightInline{Page MJ, McKenzie JE, Bossuyt PM, Boutron I, Hoffmann
TC, Mulrow CD, et al. The {PRISMA} 2020 statement: An updated guideline
for reporting systematic reviews. BMJ. 2021 Mar;372:n71. }

\leavevmode\vadjust pre{\hypertarget{ref-Ivaldi2024-jl}{}}%
\CSLLeftMargin{2. }%
\CSLRightInline{Ivaldi D, Burgos M, Oltra G, Liquitay CE, Garegnani L.
Adherence to {PRISMA} 2020 statement assessed through the expanded
checklist in systematic reviews of interventions: A meta-epidemiological
study. Cochrane Evid Synth Methods. 2024 May;2(5):e12074. }

\leavevmode\vadjust pre{\hypertarget{ref-Suda2025-nd}{}}%
\CSLLeftMargin{3. }%
\CSLRightInline{Suda C, Yamamoto N, Tsuge T, Hayashi M, Suzuki K, Ikuta
Y, et al. Enhancing reporting quality using the preferred reporting
items for systematic review and meta-analysis 2020 in systematic reviews
of emergency medicine journals: A cross-sectional study. Cureus. 2025
Jan;17(1):e78255. }

\leavevmode\vadjust pre{\hypertarget{ref-Tsuge2025-ev}{}}%
\CSLLeftMargin{4. }%
\CSLRightInline{Tsuge T, Yamamoto N, Tomita Y, Hagiyama A, Shiratsuchi
D, Kato Y, et al. Reporting and methodological qualities of systematic
reviews in rehabilitation journals after 2020: A cross-sectional
meta-epidemiological study. Phys Ther. 2025 Apr;105(4). }

\leavevmode\vadjust pre{\hypertarget{ref-Kovanis2016-zy}{}}%
\CSLLeftMargin{5. }%
\CSLRightInline{Kovanis M, Porcher R, Ravaud P, Trinquart L. The global
burden of journal peer review in the biomedical literature: Strong
imbalance in the collective enterprise. PLoS One. 2016
Nov;11(11):e0166387. }

\leavevmode\vadjust pre{\hypertarget{ref-Aczel2021-hm}{}}%
\CSLLeftMargin{6. }%
\CSLRightInline{Aczel B, Szaszi B, Holcombe AO. A billion-dollar
donation: Estimating the cost of researchers' time spent on peer review.
Res Integr Peer Rev. 2021 Nov;6(1):14. }

\leavevmode\vadjust pre{\hypertarget{ref-Perlis2025-zo}{}}%
\CSLLeftMargin{7. }%
\CSLRightInline{Perlis RH, Christakis DA, Bressler NM, Öngür D,
Kendall-Taylor J, Flanagin A, et al. Artificial intelligence in peer
review. JAMA. 2025 Nov;334(17). }

\leavevmode\vadjust pre{\hypertarget{ref-Sanmarchi2023-fu}{}}%
\CSLLeftMargin{8. }%
\CSLRightInline{Sanmarchi F, Bucci A, Nuzzolese AG, Carullo G, Toscano
F, Nante N, et al. A step-by-step researcher's guide to the use of an
{AI}-based transformer in epidemiology: An exploratory analysis of
{ChatGPT} using the {STROBE} checklist for observational studies. Z
Gesundh Wiss. 2023 May;32(9):1--36. }

\leavevmode\vadjust pre{\hypertarget{ref-Alharbi2024-ar}{}}%
\CSLLeftMargin{9. }%
\CSLRightInline{Alharbi F, Asiri S. Automated assessment of reporting
completeness in orthodontic research using {LLMs}: An observational
study. Appl Sci (Basel). 2024 Nov;14(22):10323. }

\leavevmode\vadjust pre{\hypertarget{ref-Chen2025-vc}{}}%
\CSLLeftMargin{10. }%
\CSLRightInline{Chen D, Li P, Khoshkish E, Lee S, Ning T, Tahir U, et
al. {AutoReporter}: Development of an artificial intelligence tool for
automated assessment of research reporting guideline adherence. medRxiv.
2025 Apr;2025.04.18.25326076. }

\leavevmode\vadjust pre{\hypertarget{ref-Wrightson2025-pd}{}}%
\CSLLeftMargin{11. }%
\CSLRightInline{Wrightson JG, Blazey P, Moher D, Khan KM, Ardern CL.
{GPT} for {RCTs}? Using {AI} to determine adherence to clinical trial
reporting guidelines. BMJ Open. 2025 Mar;15(3):e088735. }

\leavevmode\vadjust pre{\hypertarget{ref-Srinivasan2025-gr}{}}%
\CSLLeftMargin{12. }%
\CSLRightInline{Srinivasan A, Berkowitz J, Friedrich NA, Kivelson S,
Tatonetti NP. Large language model analysis of reporting quality of
randomized clinical trial articles: A systematic review: A systematic
review. JAMA Netw Open. 2025 Aug;8(8):e2529418. }

\leavevmode\vadjust pre{\hypertarget{ref-Forero2025-ra}{}}%
\CSLLeftMargin{13. }%
\CSLRightInline{Forero DA, Abreu SE, Tovar BE, Oermann MH. Large
language models and the analyses of adherence to reporting guidelines in
systematic reviews and overviews of reviews ({PRISMA} 2020 and {PRIOR}).
J Med Syst. 2025 Jun;49(1):80. }

\leavevmode\vadjust pre{\hypertarget{ref-He2024-do}{}}%
\CSLLeftMargin{14. }%
\CSLRightInline{He J, Rungta M, Koleczek D, Sekhon A, Wang FX, Hasan S.
Does prompt formatting have any impact on {LLM} performance? arXiv
{[}csCL{]}. 2024 Nov; }

\leavevmode\vadjust pre{\hypertarget{ref-Ni2025-em}{}}%
\CSLLeftMargin{15. }%
\CSLRightInline{Ni S, Chen G, Li S, Chen X, Li S, Wang B, et al. A
survey on large language model benchmarks. arXiv {[}csCL{]}. 2025 Aug; }

\leavevmode\vadjust pre{\hypertarget{ref-Gallifant2025-nm}{}}%
\CSLLeftMargin{16. }%
\CSLRightInline{Gallifant J, Afshar M, Ameen S, Aphinyanaphongs Y, Chen
S, Cacciamani G, et al. The {TRIPOD}-{LLM} reporting guideline for
studies using large language models. Nat Med. 2025 Jan;31(1):60--9. }

\leavevmode\vadjust pre{\hypertarget{ref-noauthor_undated-xy}{}}%
\CSLLeftMargin{17. }%
\CSLRightInline{Chatbots, {ChatGPT}, and scholarly manuscripts.
\url{https://wame.org/page3.php?id=106}; }

\leavevmode\vadjust pre{\hypertarget{ref-Ross2024-hy}{}}%
\CSLLeftMargin{18. }%
\CSLRightInline{Ross R, Pillitteri V. Protecting controlled unclassified
information in nonfederal systems and organizations. Gaithersburg, MD:
National Institute of Standards; Technology; National Institute of
Standards; Technology (U.S.); 2024 May. Report No.: NIST Special
Publication (SP) 800-171 Rev. 3. }

\leavevmode\vadjust pre{\hypertarget{ref-UnknownUnknown-cn}{}}%
\CSLLeftMargin{19. }%
\CSLRightInline{Markdown.
\url{https://daringfireball.net/projects/markdown/}; }

\leavevmode\vadjust pre{\hypertarget{ref-Jacovi2023-kw}{}}%
\CSLLeftMargin{20. }%
\CSLRightInline{Jacovi A, Caciularu A, Goldman O, Goldberg Y. Stop
uploading test data in plain text: Practical strategies for mitigating
data contamination by evaluation benchmarks. arXiv {[}csCL{]}. 2023 May;
}

\leavevmode\vadjust pre{\hypertarget{ref-Li2024-xz}{}}%
\CSLLeftMargin{21. }%
\CSLRightInline{Li H, Dong Q, Chen J, Su H, Zhou Y, Ai Q, et al.
{LLMs}-as-judges: A comprehensive survey on {LLM}-based evaluation
methods. arXiv {[}csCL{]}. 2024 Dec; }

\leavevmode\vadjust pre{\hypertarget{ref-UnknownUnknown-kd}{}}%
\CSLLeftMargin{22. }%
\CSLRightInline{{ChatGPT} - {PRISMA} checker {GPTs}.
\url{https://chatgpt.com/g/g-690d3c3487988191ac6a4870545c4bd3-prisma-checker-gpts};
}

\leavevmode\vadjust pre{\hypertarget{ref-adobeextract-ya}{}}%
\CSLLeftMargin{23. }%
\CSLRightInline{{PDF} extract {API}.
\url{https://developer.adobe.com/document-services/docs/overview/pdf-extract-api/};
}

\leavevmode\vadjust pre{\hypertarget{ref-Liu2025-ky}{}}%
\CSLLeftMargin{24. }%
\CSLRightInline{Liu J, Lai H, Zhao W, Huang J, Xia D, Liu H, et al.
{AI}-driven evidence synthesis: Data extraction of randomized controlled
trials with large language models. Int J Surg. 2025 Mar;111(3):2722--6.
}

\leavevmode\vadjust pre{\hypertarget{ref-Gartlehner2024-ta}{}}%
\CSLLeftMargin{25. }%
\CSLRightInline{Gartlehner G, Kahwati L, Hilscher R, Thomas I, Kugley S,
Crotty K, et al. Data extraction for evidence synthesis using a large
language model: A proof-of-concept study. Res Synth Methods. 2024
Jul;15(4):576--89. }

\end{CSLReferences}

\clearpage
\section*{Supplemental Material}
\hypertarget{supplemental-table-1-tripod-llm-reporting-guideline-checklist}{%
\section{Supplemental Table 1 \textbar{} TRIPOD-LLM reporting guideline
checklist}\label{supplemental-table-1-tripod-llm-reporting-guideline-checklist}}

This study reports LLM evaluation of PRISMA 2020 guideline adherence
following the TRIPOD-LLM statement {[}@Gallifant2025-nm{]}. Research
design: LLM evaluation (E). LLM task: Classification (C).

\begin{longtable}[]{@{}
  >{\raggedright\arraybackslash}p{(\columnwidth - 6\tabcolsep) * \real{0.2500}}
  >{\raggedright\arraybackslash}p{(\columnwidth - 6\tabcolsep) * \real{0.2500}}
  >{\raggedright\arraybackslash}p{(\columnwidth - 6\tabcolsep) * \real{0.2500}}
  >{\raggedright\arraybackslash}p{(\columnwidth - 6\tabcolsep) * \real{0.2500}}@{}}
\toprule\noalign{}
\begin{minipage}[b]{\linewidth}\raggedright
Section
\end{minipage} & \begin{minipage}[b]{\linewidth}\raggedright
Item
\end{minipage} & \begin{minipage}[b]{\linewidth}\raggedright
Checklist item
\end{minipage} & \begin{minipage}[b]{\linewidth}\raggedright
Location in manuscript
\end{minipage} \\
\midrule\noalign{}
\endhead
\bottomrule\noalign{}
\endlastfoot
\textbf{Title} & 1 & Identify the study as developing, fine-tuning
and/or evaluating the performance of an LLM, specifying the task, the
target population and the outcome to be predicted. & Title \\
\textbf{Abstract} & 2 & See TRIPOD-LLM for abstracts (items 2a--2l
below) & Abstract \\
\textbf{Abstract: Research design} & 2a & Identify the study as
developing, fine-tuning and/or evaluating the performance of an LLM,
specifying the task, the target population and the outcome to be
predicted. & Abstract \\
\textbf{Abstract: Background} & 2b & Provide a brief explanation of the
healthcare context, use case and rationale for developing or evaluating
the performance of an LLM. & Abstract \\
\textbf{Abstract: Objectives} & 2c & Specify the study objectives,
including whether the study describes LLMs development, tuning and/or
evaluation. & Abstract \\
\textbf{Abstract: Methods} & 2d & Describe the key elements of the study
setting. & Abstract \\
\textbf{Abstract: Methods} & 2e & Detail all data used in the study,
specify data splits and any selective use of data. & Abstract \\
\textbf{Abstract: Methods} & 2f & Specify the name and version of LLM(s)
used. & Abstract \\
\textbf{Abstract: Methods} & 2g & Briefly summarize the LLM-building
steps, including any fine-tuning, reward modeling and RLHF. & N/A (no
fine-tuning performed) \\
\textbf{Abstract: Methods} & 2h & Describe the specific tasks performed
by the LLMs (for example, medical QA, summarization and extraction),
highlighting key inputs and outputs used in the final LLM. & Abstract \\
\textbf{Abstract: Methods} & 2i & Specify the evaluation
datasets/populations used, including the endpoint evaluated, and detail
whether this information was held out during training/tuning where
relevant and what measure(s) were used to evaluate LLM performance. &
Abstract \\
\textbf{Abstract: Results} & 2j & Give an overall report and
interpretation of the main results. & Abstract \\
\textbf{Abstract: Discussion} & 2k & Explicitly state any broader
implications or concerns that have arisen in light of these results. &
Abstract \\
\textbf{Abstract: Other} & 2l & Give the registration number and name of
the registry or repository (if relevant). & Data and code availability
(GitHub repository stated) \\
\textbf{Introduction: Background} & 3a & Explain the healthcare
context/use case (for example, administrative, diagnostic, therapeutic
and clinical workflow) and rationale for developing or evaluating the
LLM, including references to existing approaches and models. &
Introduction, para 1--3 \\
\textbf{Introduction: Background} & 3b & Describe the target population
and the intended use of the LLM in the context of the care pathway,
including its intended users in current gold standard practices (for
example, healthcare professionals, patients, public or administrators).
& Introduction, para 3; Discussion, para 3 \\
\textbf{Introduction: Objectives} & 4 & Specify the study objectives,
including whether the study describes the initial development,
fine-tuning or validation of an LLM (or multiple stages). &
Introduction, para 3 \\
\textbf{Methods: Data} & 5a & Describe the sources of data separately
for the training, tuning and/or evaluation datasets and the rationale
for using these data (for example, web corpora, clinical research/trial
data, EHR data or unknown). & Methods: Study design and datasets;
Methods: Data preparation and licensing \\
\textbf{Methods: Data} & 5b & Describe the relevant data points and
provide a quantitative and qualitative description of their distribution
and other relevant descriptors of the dataset (for example, source,
languages and countries of origin). & Results: Dataset characteristics;
Methods: Study design and datasets \\
\textbf{Methods: Data} & 5c & Specifically state the date of the oldest
and newest item of text used in the development process (training,
fine-tuning and reward modeling) and the evaluation datasets. & N/A
(pre-trained models used as-is; no additional training performed) \\
\textbf{Methods: Data} & 5d & Describe any data preprocessing and
quality checking, including whether this was similar across text
corpora, institutions and relevant sociodemographic groups. & Methods:
Data preparation and licensing (full-text acquisition and conversion) \\
\textbf{Methods: Data} & 5e & Describe how missing and imbalanced data
were handled and provide reasons for omitting any data. & Methods:
Statistical analysis (denominator definition); Results: Item-level error
profile (CC vs non-CC comparison) \\
\textbf{Methods: Analytical methods} & 6a & Report the LLM name, version
and last date of training. & Methods: LLM selection and locked
parameters; Table 2; Table 3 \\
\textbf{Methods: Analytical methods} & 6b & Report details of the LLM
development process, such as LLM architecture, training, fine-tuning
procedures and alignment strategy (for example, reinforcement learning
and direct preference optimization) and alignment goals (for example,
helpfulness, honesty and harmlessness). & N/A (pre-trained models used
as-is; no fine-tuning or additional alignment performed) \\
\textbf{Methods: Analytical methods} & 6c & Report details of how the
text was generated using the LLM, including any prompt engineering
(including consistency of outputs), and inference settings (for example,
seed, temperature, max token length and penalties), as relevant. &
Methods: Parameter optimization phase; Methods: LLM selection and locked
parameters; Methods: Structured outputs \\
\textbf{Methods: Analytical methods} & 6d & Specify the initial and
postprocessed output of the LLM (for example, probabilities,
classification and unstructured text). & Methods: Structured outputs \\
\textbf{Methods: Analytical methods} & 6e & Provide details and
rationale for any classification and, if applicable, how the
probabilities were determined and thresholds identified. & Methods:
Structured outputs (binary Yes/No decisions) \\
\textbf{Methods: LLM output} & 7a & Include metrics that capture the
quality of generative outputs, such as consistency, relevance, accuracy
and presence/type of errors compared to gold standards. & N/A
(classification task, not generative QA/IR/DG/SS/MT) \\
\textbf{Methods: LLM output} & 7b & Report the outcome metrics'
relevance to the downstream task at deployment time and, where
applicable, the correlation of metric to human evaluation of the text
for the intended use. & Methods: Outcomes; Methods: Statistical
analysis \\
\textbf{Methods: LLM output} & 7c & Clearly define the outcome, how the
LLM predictions were calculated (for example, formula, code, object and
API), the date of inference for closed-source LLMs and evaluation
metrics. & Methods: Outcomes; Methods: Statistical analysis; Methods:
LLM selection and locked parameters (evaluation dates: Sept--Oct
2025) \\
\textbf{Methods: LLM output} & 7d & If outcome assessment requires
subjective interpretation, describe the qualifications of the assessors,
any instructions provided, relevant information on demographics of the
assessors and interassessor agreement. & Methods: Reference standard
(two independent investigators with third-reviewer adjudication) \\
\textbf{Methods: LLM output} & 7e & Specify how performance was compared
to other LLMs, humans and other benchmarks or standards. & Methods:
Reference standard; Methods: Format and model comparison phase \\
\textbf{Methods: Annotation} & 8a & If annotation was done, report how
the text was labeled, including providing specific annotation guidelines
with examples. & Methods: Reference standard; cited source datasets
{[}@Suda2025-nd;@Tsuge2025-ev{]} \\
\textbf{Methods: Annotation} & 8b & If annotation was done, report how
many annotators labeled the dataset(s), including the proportion of data
in each dataset that was annotated by more than one annotator, and the
interannotator agreement. & Methods: Reference standard (two independent
reviewers with third adjudication) \\
\textbf{Methods: Annotation} & 8c & If annotation was done, provide
information on the background and experience of the annotators or the
characteristics of any models involved in labeling. & Methods: Reference
standard; cited source datasets {[}@Suda2025-nd;@Tsuge2025-ev{]} \\
\textbf{Methods: Prompting} & 9a & If research involved prompting LLMs,
provide details on the processes used during prompt design, curation and
selection. & Methods: Checklist formats; Methods: Parameter optimization
phase \\
\textbf{Methods: Prompting} & 9b & If research involved prompting LLMs,
report what data were used to develop the prompts. & Methods: Checklist
formats (PRISMA 2020 checklist conversion) \\
\textbf{Methods: Summarization} & 10 & Describe any preprocessing of the
data before summarization. & N/A (classification task, not
summarization) \\
\textbf{Methods: Instruction tuning/alignment} & 11 & If instruction
tuning/alignment strategies were used, what were the instructions, data
and interface used for evaluation, and what were the characteristics of
the populations doing the evaluation? & N/A (pre-trained models used
as-is; no instruction tuning performed) \\
\textbf{Methods: Compute} & 12 & Report compute, or proxies thereof (for
example, time on what and how many machines, cost on what and how many
machines, inference time, floating-point operations per second),
required to carry out methods. & Table 2; Table 3 (mean time per SR, API
cost per SR reported) \\
\textbf{Methods: Ethical approval} & 13 & Name the institutional
research board or ethics committee that approved the study and describe
the participant-informed consent or the ethics committee waiver of
informed consent. & Ethical consideration (ethics approval not required;
publicly available published articles only) \\
\textbf{Methods: Open science} & 14a & Give the source of funding and
the role of the funders for the present study. & Acknowledgements \\
\textbf{Methods: Open science} & 14b & Declare any conflicts of interest
and financial disclosures for all authors. & Competing interests \\
\textbf{Methods: Open science} & 14c & Indicate where the study protocol
can be accessed or state that a protocol was not prepared. & N/A
(protocol not pre-registered) \\
\textbf{Methods: Open science} & 14d & Provide registration information
for the study, including register name and registration number, or state
that the study was not registered. & Introduction, final para; Abstract
(GitHub repository; public artifacts upon acceptance) \\
\textbf{Methods: Open science} & 14e & Provide details of the
availability of the study data. & Data and code availability \\
\textbf{Methods: Open science} & 14f & Provide details of the
availability of the code to reproduce the study results. & Data and code
availability \\
\textbf{Methods: Public involvement} & 15 & Provide details of any
patient and public involvement during the design, conduct, reporting,
interpretation or dissemination of the study or state no involvement. &
N/A (no patient/public involvement; analysis of published systematic
reviews) \\
\textbf{Results: Participants} & 16a & When using patient/EHR data,
describe the flow of text/EHR/patient data through the study, including
the number of documents/questions/participants with and without the
outcome/label and follow-up time as applicable. & N/A (analysis of
published systematic reviews, not patient/EHR data) \\
\textbf{Results: Participants} & 16b & When using patient/EHR data,
report the characteristics overall and for each data source or setting
and development/evaluation splits, including the key dates, key
characteristics and sample size. & N/A (analysis of published systematic
reviews, not patient/EHR data) \\
\textbf{Results: Participants} & 16c & For LLM evaluation that includes
clinical outcomes, show a comparison of the distribution of important
clinical variables that may be associated with the outcome between
development and evaluation data, if available. & N/A (guideline
adherence evaluation, not clinical outcome prediction) \\
\textbf{Results: Participants} & 16d & When using patient/EHR data,
specify the number of participants and outcome events in each analysis
(for example, for LLM development, hyperparameter tuning and LLM
evaluation). & N/A (analysis of published systematic reviews, not
patient/EHR data) \\
\textbf{Results: Performance} & 17 & Report LLM performance according to
prespecified metrics (see item 7a) and/or human evaluation (see item
7d). & Results: Parameter tuning; Results: Format comparison; Results:
Model comparison; Results: Validation with rehabilitation dataset; Table
1; Table 2; Table 3 \\
\textbf{Results: LLM updating} & 18 & If applicable, report the results
from any LLM updating, including the updated LLM and subsequent
performance. & N/A (no model updating performed) \\
\textbf{Discussion: Interpretation} & 19a & Give an overall
interpretation of the main results, including issues of fairness in the
context of the objectives and previous studies. & Discussion, para
1--4 \\
\textbf{Discussion: Limitations} & 19b & Discuss any limitations of the
study and their effects on any biases, statistical uncertainty and
generalizability. & Discussion, para 5 (three limitations discussed) \\
\textbf{Discussion: Usability of the LLM in context} & 19c & Describe
any known challenges in using data for the specified task and domain
context with reference to representation, missingness, harmonization and
bias. & Discussion, para 5 (limitation 2: human annotation
inconsistencies) \\
\textbf{Discussion: Usability of the LLM in context} & 19d & Define the
intended use for the implementation under evaluation, including the
intended input, end-user and level of autonomy/human oversight. &
Discussion, para 3 (human-in-the-loop framework; model selection
guidance per use case) \\
\textbf{Discussion: Usability of the LLM in context} & 19e & If
applicable, describe how poor quality or unavailable input data should
be assessed and handled when implementing the LLM; that is, what is the
usability of the LLM in the context of current clinical care. &
Discussion, para 3 (false negative-prone items require human attention
during review) \\
\textbf{Discussion: Usability of the LLM in context} & 19f & If
applicable, specify whether users will be required to interact in the
handling of the input data or use of the LLM, and what level of
expertise is required of users. & Discussion, para 3 (reviewers verify
structured rationales alongside binary decisions; expert verification
for false positives) \\
\textbf{Discussion: Usability of the LLM in context} & 19g & Discuss any
next steps for future research, with a specific view of the
applicability and generalizability of the LLM. & Discussion, para 5
(future work: ensemble approaches, advanced reasoning, iterative
prompting); Discussion, para 6 (need for public datasets, systematic
model evaluation, hybrid workflows) \\
\end{longtable}

\emph{Abbreviations}: CC, Creative Commons; EHR, electronic health
record; IR, information retrieval; LLM, large language model; N/A, not
applicable; QA, question answering; RLHF, reinforcement learning with
human feedback; SR, systematic review.

\emph{Notes}: This study evaluated pre-trained LLMs on PRISMA 2020
guideline adherence classification without additional training,
fine-tuning, or alignment procedures (items 2g, 5c, 6b, 11, 18 not
applicable). The task is binary classification (Yes/No for checklist
items), not generative text production (item 7a, 10 not applicable).
Analysis focused on published systematic reviews rather than
patient-level data (items 16a--16d not applicable).

\end{document}